# Thermoelectric studies of electronic properties of ferromagnetic GaMnAs layers

V. Osinniy, K. Dybko, A. Jedrzejczak, M. Arciszewska, W. Dobrowolski, and T. Story
*Institute of Physics, Polish Academy of Sciences, Al. Lotników 32/46, 02-668 Warsaw, Poland*

M.V. Radchenko, V.I. Sichkovskiy, and G.V. Lashkarev
*Institute of Problems of Material Science, National Ukrainian Academy of Sciences
Krzizhanovskogo Str. 3, 252180 Kiev, Ukraine*

S.M. Olsthoorn
*Research Institute for Materials, High Magnetic Field Laboratory, University of Nijmegen,
Toernooiveld 1, 6525 ED Nijmegen, The Netherlands*

J. Sadowski
*Max-lab, Lund University, SE-221 00 Lund, Sweden, and
Institute of Physics, Polish Academy of Sciences, Al. Lotników 32/46, 02-668 Warsaw, Poland*

Thermoelectric power, electrical conductivity, and high field Hall effect were studied over a broad temperature range in ferromagnetic $Ga_{1-x}Mn_xAs$ epitaxial layers ($0.015 \leq x \leq 0.06$). Thermoelectric power analysis gives the information about carrier transport mechanisms in layers with both metallic and non-metallic type of conductivity and allows determination of the Fermi energy and carrier concentration. At high temperatures (T>70 K) the thermoelectric power in GaMnAs linearly increases with increasing temperature. That indicates the presence of a degenerate hole gas with the Fermi energy $E_F$=220±25 meV, nearly independent of Mn content (for $0.02 \leq x \leq 0.05$). At lower temperatures GaMnAs layers with metallic-type conductivity show an additional contribution to the thermoelectric power with the maximum close to the Curie temperature. The layers exhibiting insulating electrical properties show 1/T-type increase of thermoelectric power at low temperatures.

## 1. Introduction

Among magnetic semiconductors suitable for future spintronic applications the ferromagnetic layers of GaMnAs remain the most important material system [1-5]. Ferromagnetic state in GaMnAs is induced by extremely high concentration of conducting holes, as shown in Zener or Ruderman-Kittel-Kasuya-Yosida interaction mean field theoretical models of ferromagnetic transition [5-8]. The electronic parameters of the system, such as Fermi wavevector, carrier effective mass and concentration are the physical factors determining not only the transition temperature but also the magnetic state of the system. An example of the direct influence of electronic parameters of GaMnAs on its magnetic properties is the control of the direction of the easy axis of magnetizaton achieved for various concentrations of conducting holes [9-11]. Ferromagnetic GaMnAs layers are grown by low-temperature molecular beam epitaxy (LT-MBE) technique, which allows to obtain monocrystalline single phase layers with Mn content up to 9 at. % [1,12-15]. These layers constitute a system which is strongly disordered in both magnetic (random site diluted magnetic system) and electronic (very strong compensation of Mn acceptor centers by donors) subsystems. It results in extremely intriguing electronic properties of this material revealing both a semi-metallic character of conductivity (usually for $0.03 \leq x \leq 0.055$) and an insulating one observed both for higher and lower Mn content [1,5,10,16-18]. In GaMnAs layers, Mn ions serve as the source of local magnetic moments as well as electrically active acceptor centers generating very high concentration of conducting holes of the order of $10^{20}$ $cm^{-3}$. The underlying mechanism of ferromagnetic transition and magnetotransport effects is the *p-d* exchange coupling between the local magnetic moments of Mn ions and spins of conducting holes. It also is the driving mechanism for electrically or optically detected spintronic effects [1,10,19-21]. The LT-MBE GaMnAs layers are known to be strongly compensated by double-donor centers: interstitial Mn ions as well as As antisites [14,22-25]. The control over this heavy background doping achieved recently via thermal annealing procedures resulted in a substantial increase of the ferromagnetic Curie temperature, up to about 170 K [23-29]. Although intriguing and challenging scientifically as well as important for achieving a practical control of ferromagnetic transition temperature and magnetic anisotropy in GaMnAs, the electronic properties of this material are relatively poorly understood. It is related to the strong limitations for the use of standard magnetotransport and magnetooptical methods in materials with very low carrier mobilities and a heavy electrical compensation as well as a strong band tailing and other disorder driven electronic effects encountered in LT-MBE GaMnAs layers [30,31].

In this work, we examine the electronic properties of GaMnAs layers by studying the electron transport effect - the thermoelectric power (Seebeck effect). This experimental method proved to be very successful in the studies of electronic parameters of various metallic and semiconducting materials. It is particularly useful in the studies of strongly disordered electronic systems exhibiting variety of electron transport mechanisms, involving conduction or valence band states as well as impurity bands or hopping mechanism [32,33]. The analysis of the temperature dependence of the thermoelectric power allows for the determination of the Fermi energy and (within a certain model of the band structure) of carrier concentration or the effective mass of density of states. We are interested in both determining these basic electronic parameters of GaMnAs and studying the influence of the ferromagnetic transition on electronic transport effects. Therefore, our experimental studies cover a broad temperature region that includes both paramagnetic and ferromagnetic range of temperatures. For this study, we selected a set of thick (up to 3 microns), high quality monocrystalline layers of $Ga_{1-x}Mn_xAs$ grown by LT-MBE technique on GaAs (001) semi-insulating substrates with the Mn content ($0.015 \leq x \leq 0.06$) covering both the layers with metallic and non-metallic character of electrical conductivity. We show that at high temperatures (T>70 K) the temperature dependence of thermoelectric power in GaMnAs layers is of quasi-linear character expected for a strongly p-type doped degenerate semiconductor. The analysis of this standard (diffusion) contribution to the thermoelectric power allowed to evaluate the Fermi energy ($E_F$=0.22 eV) and carrier concentration (p=$2\times10^{20}$ $cm^{-3}$), which turned out to be very weakly dependent on the content of Mn ions. At lower temperatures, the layers of GaMnAs with metallic type of conductivity reveal an additional contribution to the thermoelectric power attributed to the ferromagnetic transition (Kasuya mechanism [34]). In the low, but paramagnetic, range of temperatures we also observe a negative, temperature independent, contribution to the thermoelectric power assigned to the exchange thermoelectric power (Kondo mechanism [35]). In GaMnAs layers exhibiting insulating properties the additional $\Delta\alpha \propto \Delta E/k_B T$-type contribution to thermoelectric power is observed at low temperatures with the activation energy ($\Delta E$=1 meV) characteristic for hopping conduction mechanisms.

The paper is organized as follows. In Section 2, after giving the information on the technological details of growth as well as structural, electrical, and magnetic characterization of the layers, we present the results of the experimental studies of the temperature dependence of thermoelectric power and high field Hall effect in a series of GaMnAs layers covering a broad range of Mn compositions. The analysis of the experimental observations, covering high temperature as well as low temperature (both paramagnetic and ferromagnetic) regions is given in Section 3. In this section, we also discuss the physical mechanisms likely to contribute to thermoelectric effects in GaMnAs, and perform the quantitative analysis of the experimental data on the thermoelectric power in GaMnAs at high temperatures. It results in the determination of the Fermi energy and conducting hole concentration in the layers. The paper is summarized in Section 4.

## 2. Growth, characterization, and transport measurements

The monocrystalline epitaxial layers of $Ga_{1-x}Mn_xAs$ (0.015≤x≤0.065, 0.3-3 μm thick) were grown by low-temperature molecular beam epitaxy technique on semi-insulating GaAs (001) substrates with a buffer layer of high-temperature MBE GaAs (with thickness of 0.1-0.2 μm). Due to technological regime applied during the MBE growth, the layer of GaMnAs and the GaAs buffer layer are separated by an ultrathin (with thickness of about 4 nm) n-type layer of low-temperature MBE GaAs. The GaMnAs layers were grown at the substrate temperature of 230 ℃ using effusion cells for Ga, Mn, and $As_2$ with the ratio of $As_2$ to Ga fluxes of about 2. The layers were not post-growth annealed. The list of the samples investigated in this work is presented in Table 1. Further detailed information about the growth conditions and structural characterization of the layers is given in Refs. 13 and 14. In a relatively broad temperature range discussed in this work, a certain contribution to the electrical conductivity of the GaMnAs/GaAs heterostructure is expected not only from the top layer of GaMnAs (p≈$10^{20}$ $cm^{-3}$, thickness d=0.3-3 μm) but also from the buffer layers of GaAs grown by high-temperature (p≈$10^{13}$-$10^{14}$ $cm^{-3}$, d=0.1-0.2 μm) and low-temperature MBE GaAs (n-type, high resistivity, d=4 nm) buffer. Therefore, we checked, applying formulae for conductivity and thermoelectric power of a system of parallel conducting layers, that in the structures studied in this work the conductivity and thermoelectric power contributions due to the buffer layers can be neglected [36]. This conclusion is confirmed by our experimental observations showing practically no difference between thermoelectric properties of GaMnAs layers with similar Mn content but large (up to factor of 10) difference of layer thickness.

The measurements of the magnetic susceptibility, χ, were carried out over the temperature range 4.5-100 K by applying the ac magnetic field $H_{ac}$=5 Oe at a frequency of f=625 Hz. We found that all GaMnAs layers investigated undergo ferromagnetic transition with the Curie temperature between $T_C$=30 K (sample with 1.5 at. % of Mn) and $T_C$=60 K (sample with 6 at. % of Mn). The transition is evidenced by a sharp increase of the magnetic susceptibility below a certain temperature as well as an observation of a peak in the χ(T) plot (Fig. 1). In some layers, the peak in the χ(T) dependence was observed at noticeably lower temperature than the temperature corresponding to the sharp increase of the susceptibility. In such a case we show in Fig. 2 both of these characteristic temperatures. For the GaMnAs layer with the nominal Mn concentration of 6 at. % we found two peaks in the χ(T) dependence (see Fig. 1). It may indicate that this layer is composed of two layers with Mn content of 6 at.% and about 1 at.%.

The resistivity (at zero magnetic field) was measured by the standard four probe method over the temperature range 4.2-300 K. As shown in Fig. 3, we observed both metallic and insulating character of electrical conductivity in GaMnAs layers in the low temperature limit. This effect is well known in LT-MBE GaMnAs layers. Usually, in unannealed layers, metallic character of conductivity is found in GaMnAs layers with Mn content between 3 and 5.5 at. %, whereas samples with Mn content both below 3 at. % and above 5.5 at. % behave (in the low temperature limit) as insulators. The ferromagnetic transition in GaMnAs is usually detected also electrically as a characteristic feature (a maximum or a kink) in the temperature dependence of the resistivity. The effect is believed to originate from the scattering on the thermodynamic fluctuations of the magnetic subsystem of Mn ions and is expected to have the maximum at the Curie temperature. We found similar behavior in our layers, as shown in Fig. 3 The ferromagnetic transition temperatures determined in such a way correspond well to the maxima observed experimentally on the temperature dependence of magnetic susceptibility.

The measurements of the thermoelectric power of LT-GaMnAs were carried out in the temperature range T=15-200 K by using a standard method with the temperature difference along the sample $\Delta T$=0.4-2 K and with no magnetic field applied. The temperature difference in the sample was monitored by fine Au:Fe-chromel thermocouples attached to the GaMnAs layer by conducting silver paste. It provided good thermal and electrical contact between the layer and the thermocouple wires. The typical area of the GaMnAs layers was 10x3 mm$^2$. The temperature dependence of the thermoelectric power in GaMnAs layers is presented in Fig. 4a. The behavior of all the samples in the high temperature range (above 70 K) is quasi-linear. In the low temperature range the essential difference appears for metallic and nonmetallic samples. In order to point out various types of low temperature behavior of thermoelectric power of GaMnAs we show in Fig. 4b the $\Delta\alpha$ part obtained from the experimentally measured values of thermoelectric power by subtracting a linear (aT+b) contribution based on the extrapolation from the 70-200 K region.

Electrical properties of the same layers were also studied by the high field dc Hall effect and magnetoresistance measurements in order to provide the control experiment for the comparison of the thermoelectric and the Hall effect measurements of conducting hole concentration in GaMnAs. The Hall effect was measured for a series of the samples of GaMnAs at the temperature T=1.1 K by applying a magnetic field up to 20 T. The low temperature and very high magnetic fields are essential in this experiment because one needs to saturate magnetically the layers in order to reveal the contribution of the ordinary Hall effect on a strong background of the anomalous Hall effect [16]. Although 5 layers of GaMnAs with different Mn content (x=0.015, 0.03, 0.04, 0.045, and 0.05) were experimentally studied, only for the layers with x=0.04 and 0.045 (exhibiting metallic type of conductivity) we found that the ordinary Hall effect was observed in the high magnetic field (above 15 T). The concentration of conducting holes determined from the high field Hall effect measurements is shown in Table 1 and in Fig. 5b. Considering the experimental limitations in determining the ordinary Hall effect in the presence of strong anomalous Hall effect and magnetoresistance, the agreement of the two methods is quite satisfactory.

### 3. Discussion

The measurements of the thermoelectric power in GaMnAs layers show clearly that there exist two distinct temperature regions. At high temperatures (approximately for T>70 K) all the layers studied show qualitatively similar linear increase of the thermoelectric power with increasing temperature. This is the standard behavior expected for strongly doped semiconductors with a degenerate hole gas. For the interpretation of the experimental results in this temperature region one can apply well known theoretical results for diffusion thermoelectric power. At low temperatures (T<70 K), the temperature dependence of the thermoelectric power of GaMnAs layers qualitatively deviates from the standard expectations both in paramagnetic and in ferromagnetic range of temperatures. These additional contributions to thermoelectric power, characteristic for diluted magnetic systems, are brought about by spin dependent scattering and spin splitting of the band states, as shown by the models of Kondo [35] and Kasuya [34].

### 3.1 Thermoelectric power at high temperatures

At high temperatures (T>70 K), the temperature dependence of the thermoelectric power of all GaMnAs layers studied is quasi-linear. This behavior is qualitatively similar to the temperature dependence of the thermoelectric power observed in other heavily p-type doped semiconductor alloys, e.g. in SiGe:P [37]. To analyse quantitatively the experimental data, we use the standard expression for the thermoelectric power in the heavily doped semiconductors with a parabolic energy dispersion relation. In the limit of high degeneration of electron or hole gas it reads:

$$\alpha_D = \frac{\pi^2 k_B^2 T}{3qE_F}(r + 3/2) \,, \tag{1}$$

where r is the scattering parameter, $E_F$ is the Fermi energy, $k_B$ is the Boltzmann constant, and q is the charge of conducting carriers.

To calculate the Fermi energy, the scattering parameter r corresponding to the mechanism that limits the absolute value of the carrier mobility has to be specified. In our analysis we take that this mechanism is the scattering on ionized impurities (r=3/2), the huge concentration of which (both Mn acceptors and compensating donor centers) is present in GaMnAs. The Fermi energies obtained within the approach described above are presented in Fig. 5a. For the samples with metallic type of conductivity the Fermi energy changes only slightly and it is found about 220 meV below the top of the valence band, whereas for the samples with non-metallic type of conductivity the Fermi energy value decreases by about 50 meV. Taking the effective mass of the density of states equal to $m^*=0.5m_0$ (the value frequently used in the analysis of electronic properties of GaMnAs [1,18]), we obtain the concentration of conducting holes presented in Fig. 5b. It agrees well with the high field Hall effect measurements made for these layers.

The dependence of the concentration of conducting holes on Mn content, presented in Fig. 5a, can be qualitatively explained by taking into account the presence of very high concentration of two types of compensating deep donor centers in LT-MBE GaMnAs layers. In GaMnAs it is observed that despite the presence of very high (of the order of $10^{20}$ cm$^{-3}$) concentration of Mn acceptor centers the layers with Mn content lower than about 1 at. % exhibit non-metallic type of conductivity and become highly resistive upon lowering temperature. It is due to the compensation of Mn acceptors by native defects ($As_{Ga}$ antisites), which are deep double donor centers well known in LT-MBE GaAs. Upon further increasing Mn concentration, a metallic p-type conductivity is achieved with conducting hole concentration of the order of $10^{20}$ cm$^{-3}$. For Mn content exceeding about 5 at. %, the efficiency of the p-type doping with Mn is gradually reduced. It is due to the appearance of an increasing concentration of Mn ions occupying lattice interstitial positions; in that case Mn is a double donor center. It eventually leads to a decrease of conducting hole concentration.

In our quantitative analysis we used theoretical results developed for the case of degenerate semiconductor system with a band-type conductivity mechanism. The only band structure parameter needed in this approach is the density of states effective mass. This approach is expected to be valid for parabolic, anisotropic (also multiple degenerate) bands. However, it neglects such factors as, e.g., nonparabolicity of the valence band, strain induced removal of degeneracy of the light- and heavy-hole bands, and possible contribution due to holes from the spin-orbit split valence band. This approach as well as the effective mass parameter used corresponds to the approximation frequently made, e.g., in the analysis of the ordinary Hall effect or conductivity of GaMnAs. The fully satisfactory theoretical model for the thermoelectric effects in heavily p-type doped GaMnAs should, apart from the realistic description of its complicated valence band, take also into account the modification of electronic valence band states close to the metal–insulator transition in the presence of *p-d* exchange interaction. To our knowledge, such a theoretical model is yet not available.

We found the Fermi level located rather deep below the top of the valence band, i.e. in agreement with the models of the electronic band structure of GaMnAs adopted in theoretical calculations of carrier induced ferromagnetism, spin-dependent tunnelling, and other magnetotransport and magnetooptical experiments [1,20,38,39]. We note also that the dependence of the Fermi level in GaMnAs on Mn content obtained in our work is qualitatively very similar to the dependence determined based on the photoemission experiments. However, the two results differ with respect to the absolute position of the Fermi level, which in photoemission experiments was found at (or just above) the top of the valence band [40]. In our work, the position of the Fermi level is given with respect to the top of the valence band, i.e., the edge of the energy range in which conducting hole states are available in GaAs. Although certain distortion of the density of states of GaMnAs in this range of energies is inevitable and this reference point may require some correction, we expect that the qualitative picture should not change.

We also note that the linear (aT+b) extrapolation of the temperature dependence of the thermoelectric power of GaMnAs from high temperatures to T=0 K shows a negative value of the parameter b (b=–(10-20) μV/K). Therefore, the experimental value of the diffusion contribution to the thermoelectric power should be determined rather from the slope of the linear part of α(T) plot than from the value of thermoelectric power at a given temperature [36]. This behavior found in GaMnAs is in contrast to thermoelectric power studies, e.g., of p-PbSnMnTe degenerate ferromagnetic semiconductor and other nonmagnetic semiconducting and metallic materials [41], for which the linear dependence is usually observed down to liquid helium temperatures [41]. The physical origin of this temperature independent negative contribution to the thermoelectric power of GaMnAs is likely to be the exchange thermoelectric power proposed by Kondo [35] as will be discussed in section 3.2.1.

### 3.2 Thermoelectric power at low temperatures

#### 3.2.1 Paramagnetic region

In diluted magnetic semiconducting and metallic systems with strong sp-d exchange coupling the new mechanism of the additional thermoelectric power is expected. It stems from the spin dependent scattering of carrier on magnetic ions. The full analysis of the influence of these mechanisms on the resisitivity and the thermoelectric power was given by Kondo [35]. This, so called, exchange thermoelectric power is only weakly temperature dependent and was found in Kondo systems together with the well known effect of logarithmic resistance minimum. Although, quantitatively, the theory of Kondo is valid only for the limit of extreme dilution, for systems with higher (but paramagnetic) concentration of magnetic ions the extension of this theory was proposed [35] with the following expression for the temperature dependence of the exchange thermoelectric power:

$$\alpha_{ex}(T) = \alpha_0 \frac{T}{T + T_0}, \text{ where } \alpha_0 = 4\pi^2 \frac{k_B}{q} D^2(E_F) I_{pd} V \frac{\rho_{ex}}{\rho} \tag{2}$$

Here $D(E_F)$ is the density of states at the Fermi level, $I_{pd}$ is the $p$-$d$ exchange integral between the carriers and the magnetic centres, V is the nonmagnetic scattering potential and $\rho_{ex}/\rho$ is the ratio of the exchange part to the total resistivity. $T_0$ is a material parameter that is independent of temperature and is only weakly dependent on the concentration of magnetic ions. In the limit of very low concentration of Mn ions, $T_0$ is the Kondo temperature. The estimation of the $\alpha_0$ with following parameters corresponding to GaMnAs: $D(E_F)=8*10^{21}$ eV$^{-1}$cm$^{-3}$, $I_{pd}$=-1 eV, V=1 eV, and $\rho_{ex}/\rho$=1/4, gives $\alpha_0 = -20\mu$V/K.

For the analysis of the temperature dependence of thermoelectric power in metallic GaMnAs samples in the entire (paramagnetic) range of temperatures we use the following expression:

$$\alpha = \alpha_{dif} + \alpha_{ex} = bT + \alpha_0 \frac{T}{T + T_0}, \qquad (3)$$

where the first term is due to the diffusion contribution to the thermoelectric power and the second is due to the exchange contribution.

The example of the comparison of the model calculations and the experimental data is presented in Fig. 7. The best fit was achieved for the parameters $\alpha_0$=-80 μV/K, $T_0$=150 K, which are of correct order of magnitude. As expected, the incorporation of the exchange (Kondo) contribution explains the negative offset value of the thermoelectric power discussed above as well as qualitatively reproduces the very weak temperature dependence observed at low temperatures.

### 3.2.2 Ferromagnetic region

Below the ferromagnetic transition temperature, the additional contribution to the thermoelectric power is expected due to the appearance of the exchange splitting of the band states in metals or semiconductors. This ferromagnetic contribution to the thermoelectric power, $\alpha_{FM}$, is described in the framework of Kasuya model of thermoelectric effects in metallic systems with the $sp$-$d$ exchange coupling [34]. It is based on the solution of Boltzmann equation taking into account two effects originating from the $sp$-$d$ exchange interaction between the magnetic ions and the conducting carriers: (1) the exchange splitting of the conduction band states and (2) the spin-dependent carrier scattering. In contrast to the model of Kondo, the Kasuya contribution is nonzero only below the Curie temperature. The expression for the thermoelectric power below the Curie temperature reads [34]:

$$\alpha = \alpha_{dif} + \alpha_{FM} = \frac{\pi^2 k_B^2 T}{3qE_F}(r + 3/2) - FA(w)\bar{J}_z^2(w), \qquad (4)$$

where F is a (temperature independent) material parameter depending, in particular, on the third power of the $p$-$d$ exchange constant (F~$I_{pd}^3$) and:

$$A(w) = \frac{2w^2}{e^w + e^{-w} - 2}, \quad \bar{J}_z(w) = J + \frac{2J+1}{e^{(2J+1)w} - 1} - \frac{1}{e^w - 1}, \qquad (5)$$

$$w = \frac{H_m \bar{J}_z}{k_B T} \text{ and } H_m = \frac{3k_B \Theta}{J(J+1)}.$$

Here J is the quantum number of the total angular momentum of magnetic impurity and for Mn ions in GaMnAs we take J=5/2. $H_m$ is the molecular field parameter directly proportional to the paramagnetic Curie temperature $\Theta$. Fig. 6 presents the comparison of the Kasuya model and the ferromagnetic contribution observed in GaMnAs layer A268 with 3 at. % Mn. In the calculations, the ferromagnetic Curie temperature obtained from the magnetic susceptibility measurements was used. Despite the model character of the calculation as well as some contribution to the thermoelectric power expected also above the Curie temperature, the qualitative agreement is achieved.

Finally, we would like to address the problem of the physical justification of using the additive diffusion, paramagnetic Kondo and ferromagnetic Kasuya contributions for the description of the thermoelectric power in GaMnAs layers in the entire temperature range discussed. The key point is that at low temperatures all three contributions are needed to qualitatively reproduce the experimentally observed behavior. At the same time, the temperature regions of applicability of the theoretical models of Kondo (only for paramagnetic case) and Kasuya (nonzero contribution only below $T_c$) are different. Although the theoretical model that would unify the two approaches both in the paramagnetic and ferromagnetic range of temperatures is not available, it probably could not be reduced to the simple addition of both contributions. In this respect, we would like to point out that there is a possibility that all these contributions are observed in the entire temperature range studied because of the effect of electronic separation, which produces in GaMnAs layers both ferromagnetic (high enough hole concentration) and paramagnetic (deficient hole concentration) regions with relative volume fraction varying depending on the Mn content, doping, and growth conditions. The possibility of such an effect is currently discussed, e.g., in manganites but it was originally suggested for magnetic semiconductors with relatively low carrier concentration [42]. The existence of such an effect would also explain our experimental findings that, between the limiting cases of metallic and highly resistive layers, there exists in fact a continuity of low temperature behaviors of thermoelectric power in GaMnAs. It probably reflects the changes in the relative volume of metallic and non-metallic regions of the layer.

### 3.2.3 Metallic versus non-metallic layers

Although at high temperatures there appears no qualitative difference between the temperature dependence of the thermoelectric power of metallic and nonmetallic layers of GaMnAs, their low temperature thermoelectric properties qualitatively differ. For layers exhibiting metallic type of conductivity, the low temperature behavior of thermoelectric power is described by the superposition of ferromagnetic Kasuya contribution and paramagnetic exchange thermoelectric power, as is discussed above. For layers exhibiting nonmetallic type of conductivity at low temperatures, the thermoelectric power strongly increases upon decreasing temperature and the $\Delta\alpha \sim \Delta E/k_B T$ type of behavior is found. This type of temperature dependence of thermoelectric power is observed, e.g., in systems with constant range hopping conduction [32]:

$$\alpha_{hop} = \frac{k_B}{q}\left(\frac{\Delta E}{k_B T} + A\right),$$ (6)

where $\Delta E$ is the activation energy whereas A is a parameter only slightly dependent on temperature [32]. Based on the experimentally determined slope of the $\Delta\alpha$ vs 1/T plot (see Fig. 8), one can calculate the activation energy of the electron transport mechanism. The experimentally found value $\Delta E=1$ meV suggests the hopping type of the conduction mechanism. The analysis of electrical conductivity of the same layer is less clear in this temperature range due to the nonmonotonic temperature dependence of carrier mobility.

## 4. Summary

We examined the electronic properties of thick monocrystalline layers of $Ga_{1-x}Mn_xAs$ grown by low-temperature molecular beam technique on GaAs (001) semi-insulating substrates with the Mn content ($0.015 \leq x \leq 0.06$) range covering both the layers with metallic and non-metallic character of electrical conductivity. We applied the well known electronic transport method, and the analysis of the thermoelectric power in a broad temperature region that includes both paramagnetic and ferromagnetic range of temperatures

At high temperatures (T>70K), the thermoelectric power in GaMnAs layers linearly increases with increasing temperature as expected for a strongly p-type doped degenerate semiconductor. In this temperature region the standard analysis of the diffusion contribution allows to evaluate the Fermi energy ($E_F=220\pm25$ meV) and carrier concentration ($p=2\pm0.2\times10^{20}$ cm$^{-3}$) which turned out to be very weakly dependent on the content of Mn ions. At lower temperatures (T<70 K), metallic layers of GaMnAs reveal an additional contribution attributed to the ferromagnetic transition (Kasuya mechanism). At low but paramagnetic range of temperatures we also observe a temperature independent contribution to the thermoelectric power assigned to the exchange thermoelectric power (Kondo mechanism). In GaMnAs layers exhibiting insulating properties the additional $\Delta E/k_B T$-type contribution to thermoelectric power at low temperatures is observed and attributed to hopping conduction mechanisms.

### Acknowledgements

This work was partially supported by the Committee for Scientific Research (Poland) under project PBZ-KBN-044/P03/2001 and within European Community program ICA1-CT-2000-70018 (Center of Excellence CELDIS). The GaMnAs samples were grown within the project supported by the Swedish Research Council (VR) and Swedish Foundation of Strategic Research (SSF).

**Table 1** Structural, magnetic, and electrical parameters of GaMnAs layers.

x - molar fraction of Mn; d - thickness of GaMnAs layer; $T_C$ (susc.) – ferromagnetic Curie temperature determined from magnetic susceptibility measurements; $T_C$ (res.) - ferromagnetic Curie temperature determined from resistivity measurements; $E_F$ - Fermi energy at 120 K; $p_{TP}$ - hole concentration determined from the thermoelectric power measurements (at 120 K); and $p_H$ - hole concentration determined from high field Hall effect measurements (at 1.1 K).

| sample | x | d $\mu m$ | $T_C$ (susc.) K | $T_C$ (res.) K | $E_F$ meV | $p_{TP}$ $10^{20}$ cm$^{-3}$ | $p_H$ $10^{20}$ cm$^{-3}$ |
|--------|-------|-----|-----------|----------|-----|------|-----|
| A203 | 0.015 | 0.3 | 16 | 26 | 170 | 1.25 | - |
| A64 | 0.02 | 0.2 | 36 | 37 | 240 | 2.6 | - |
| A268 | 0.03 | 1 | 46 | 48 | 230 | 2.5 | - |
| A377 | 0.04 | 3 | 40 | 42 | 205 | 2.2 | 3.5 |
| A204 | 0.045 | 0.3 | 38 | 32 | 230 | 2.4 | 1.6 |
| A231 | 0.05 | 0.3 | 45 | 58 | 200 | 1.9 | - |
| A266 | 0.06 | 0.3 | 12/60[+] | 18 | 160 | 1.2 | - |

[+] two peaks found on the magnetic susceptibility versus temperature plot

**Figure captions**

Fig. 1. Temperature dependence of ac magnetic susceptibility of GaMnAs layers.

Fig. 2. Ferromagnetic Curie temperature of $Ga_{1-x}Mn_xAs$ layers determined from the measurements of the temperature dependence of magnetic susceptibility (open symbols) and resistivity (full squares). For comparison a reference relation $T_C=2000*x$ is also shown as a solid line[1].

Fig. 3. Temperature dependence of zero field resistivity of GaMnAs layers.

Fig. 4. Temperature dependence of thermoelectric power of GaMnAs layers (Fig. 4a). Fig. 4b presents the low temperature behavior of thermoelectric power after subtracting the linear contribution extrapolated from the 70-200 K region.

Fig. 5. Fermi energy and hole concentration in $Ga_{1-x}Mn_xAs$ layers determined from the analysis of thermoelectric power (open dots and full squares correspond to two experimental set-ups employed) and high field Hall effect (full triangles). In the quantitative analysis of the thermoelectric measurements the scattering on ionized impurities was taken as the dominant carrier scattering mechanism with the effective mass of density of states of holes $m^*=0.5m_0$.

Fig. 6. Low temperature behavior of thermoelectric power in GaMnAs layer (A268) exhibiting metallic conductivity. The solid line presents the contribution expected from the model of Kasuya.

Fig. 7. Temperature dependence of thermoelectric power in GaMnAs layer in a broad temperature range. The lines show predictions of theoretical models involving diffusion, paramagnetic Kondo, and ferromagnetic Kasuya contributions.

Fig. 8. Temperature dependence of additional low temperature contribution to the thermoelectric power in a GaMnAs layer (A203) exhibiting non-metallic conductivity.

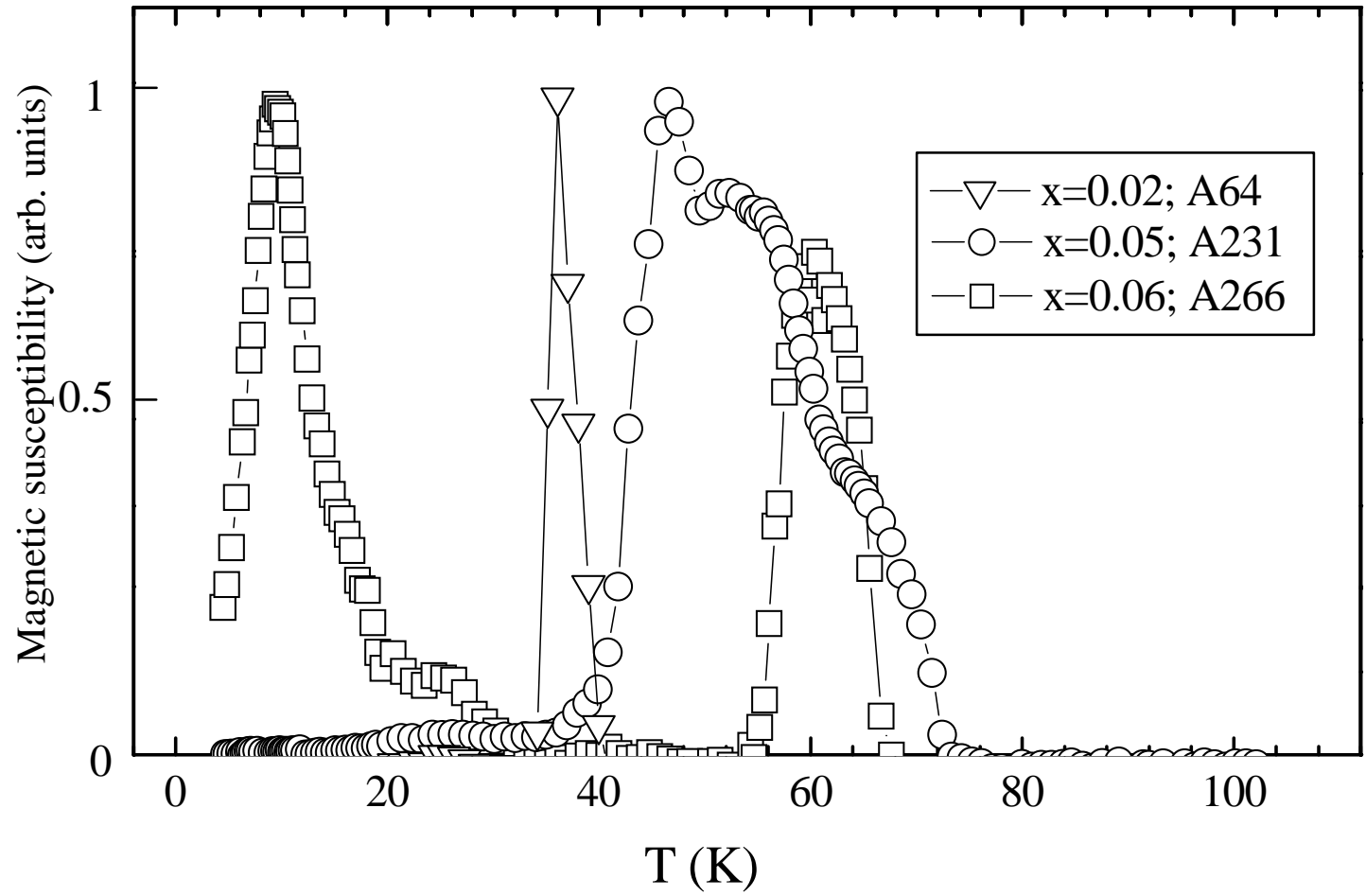

V. Osinniy et. al. Fig.1.

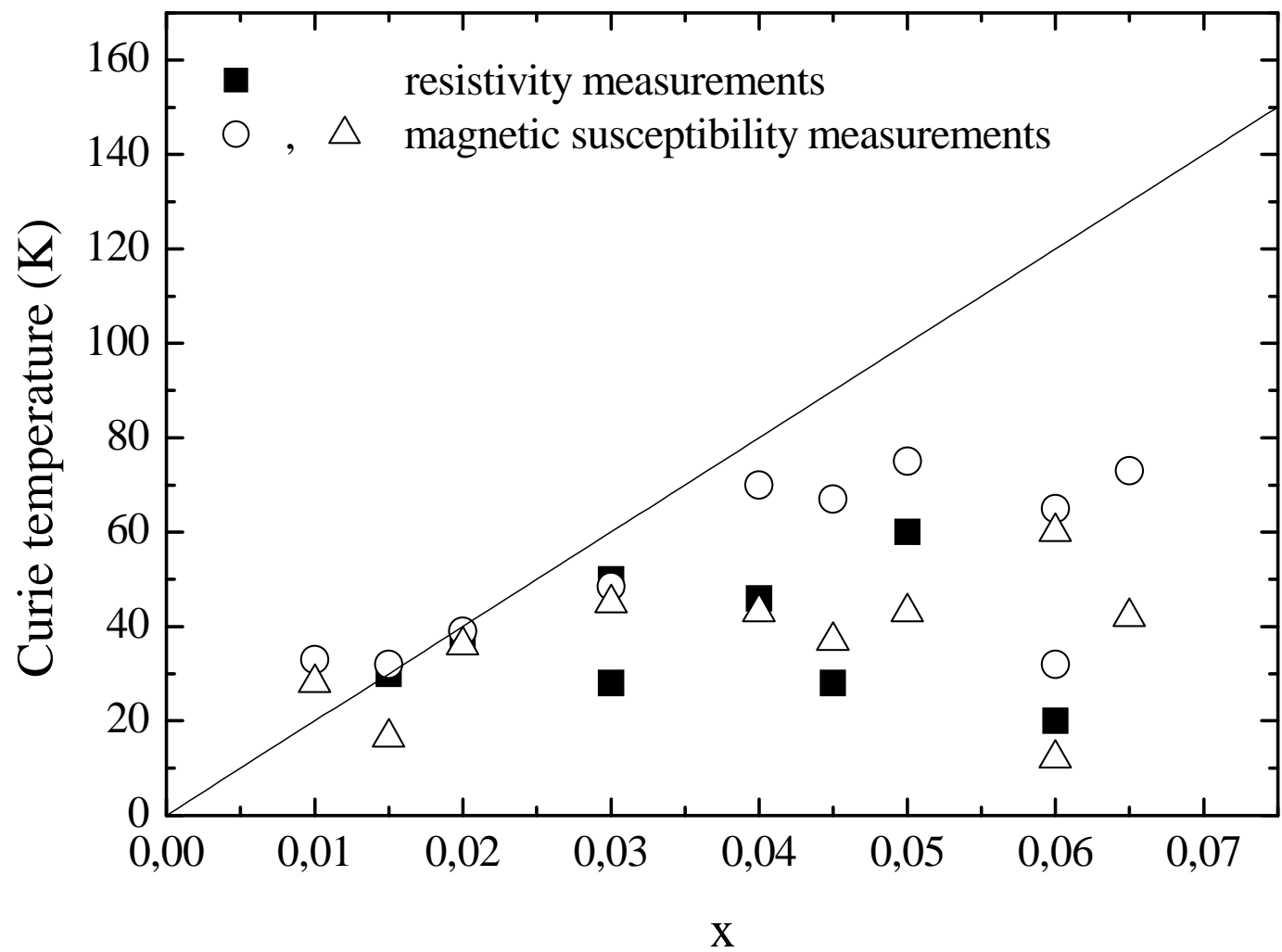

V. Osinniy et. al. Fig.2.

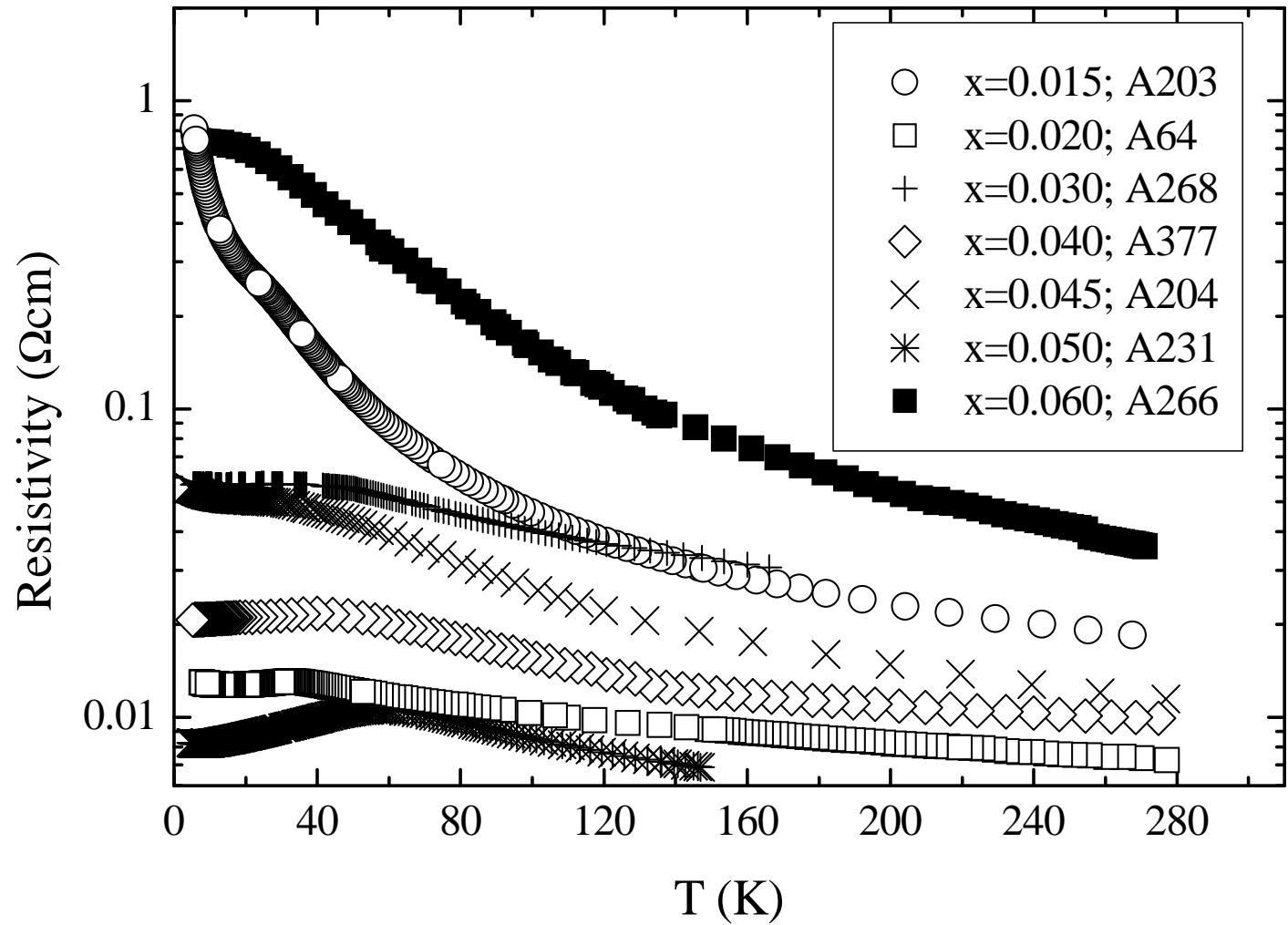

V. Osinniy et. al. Fig.3.

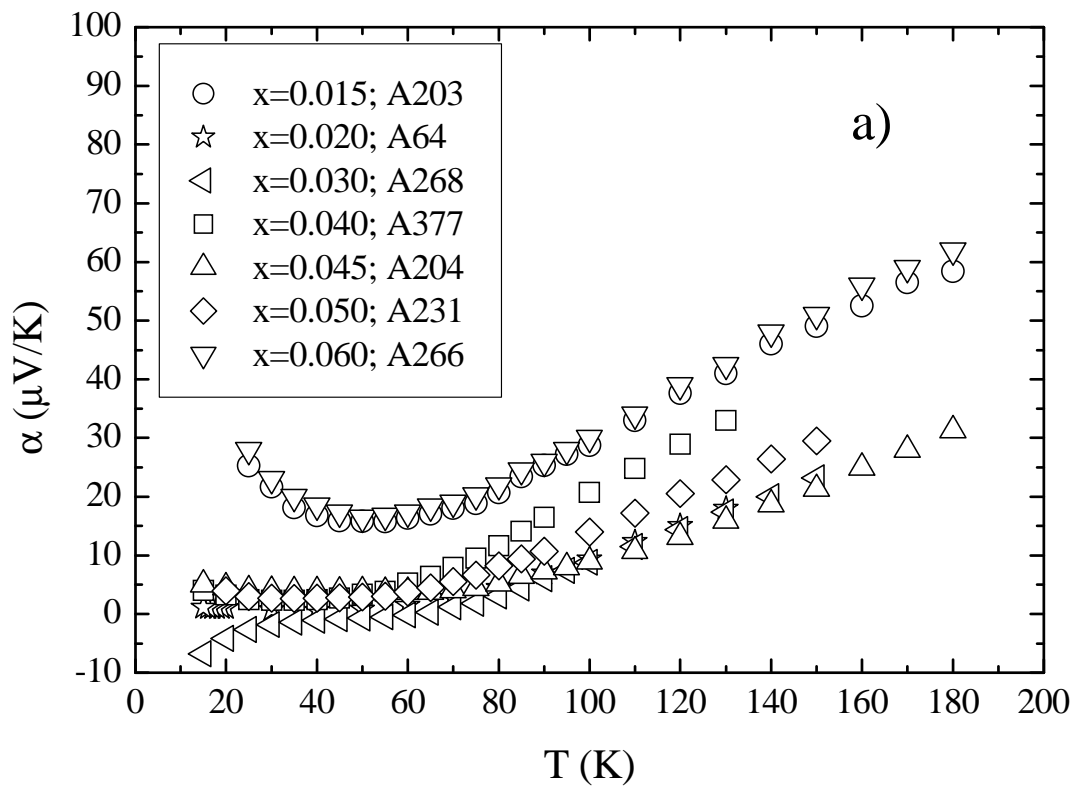

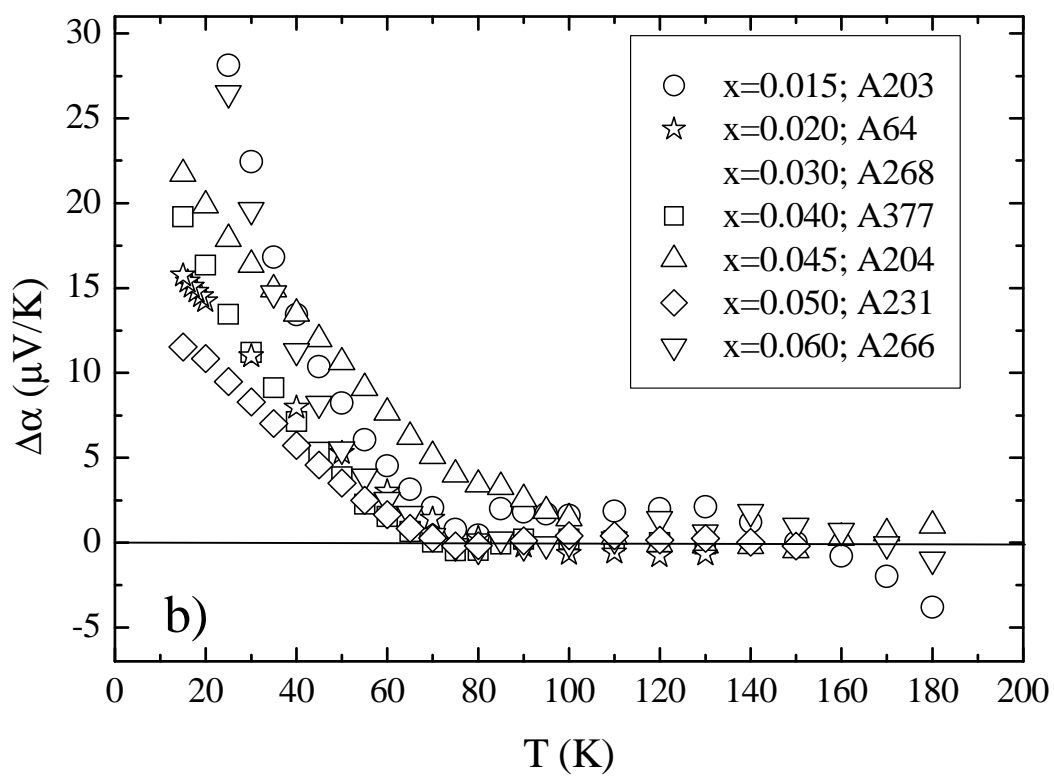

V. Osinniy et. al. Fig.4.

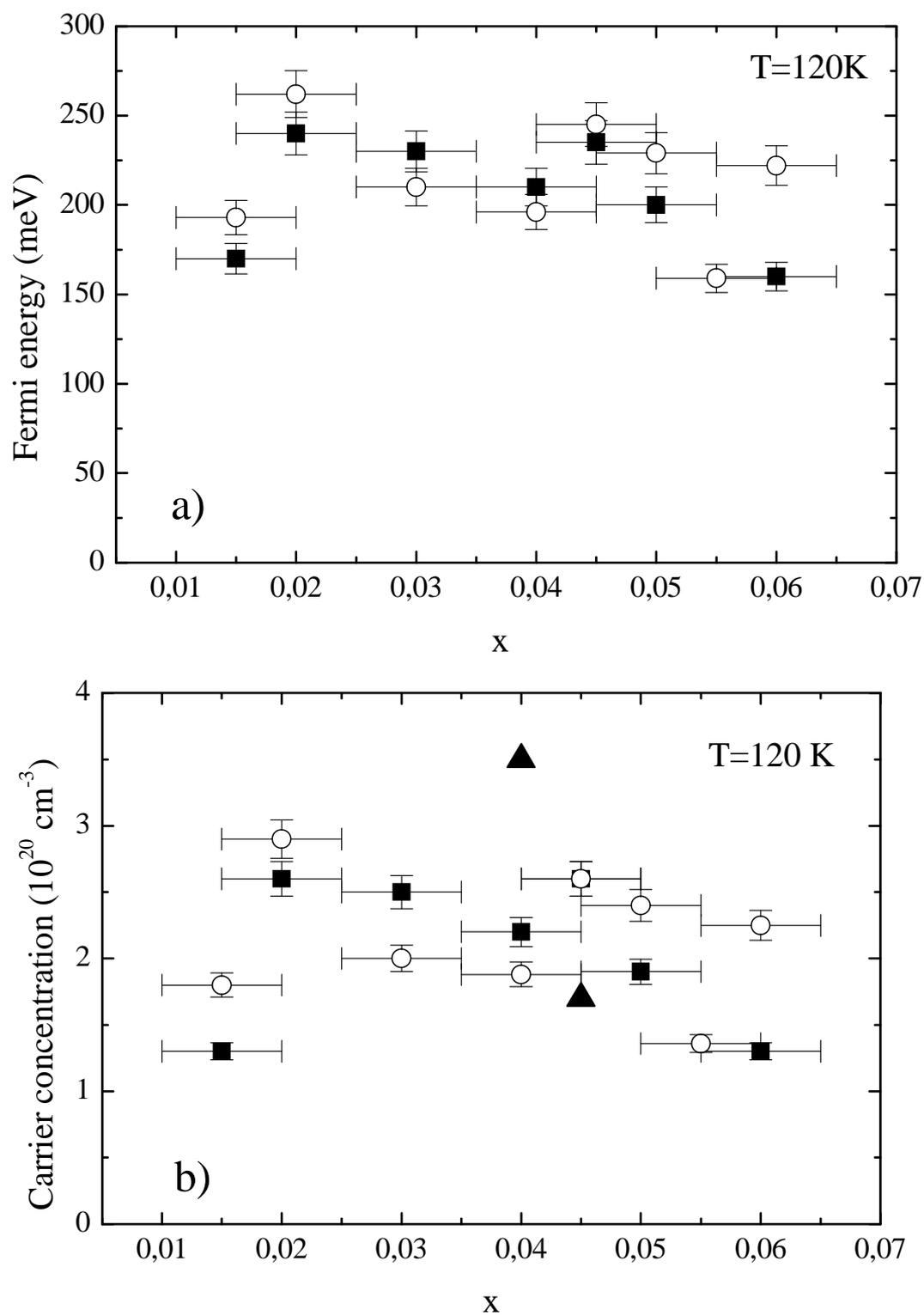

V. Osinniy et. al. Fig.5.

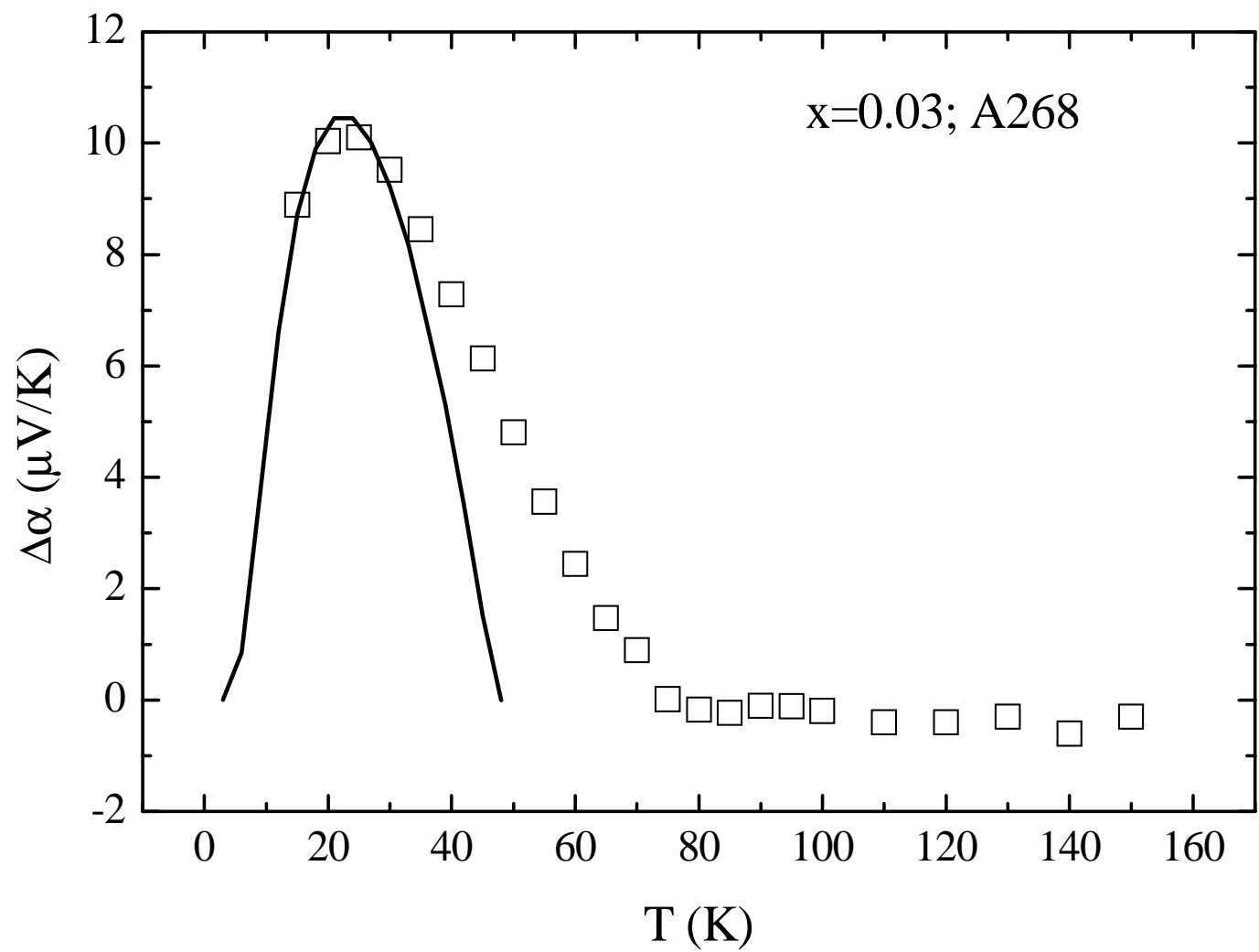

x=0.03; A268

V. Osinniy et. al. Fig.6.

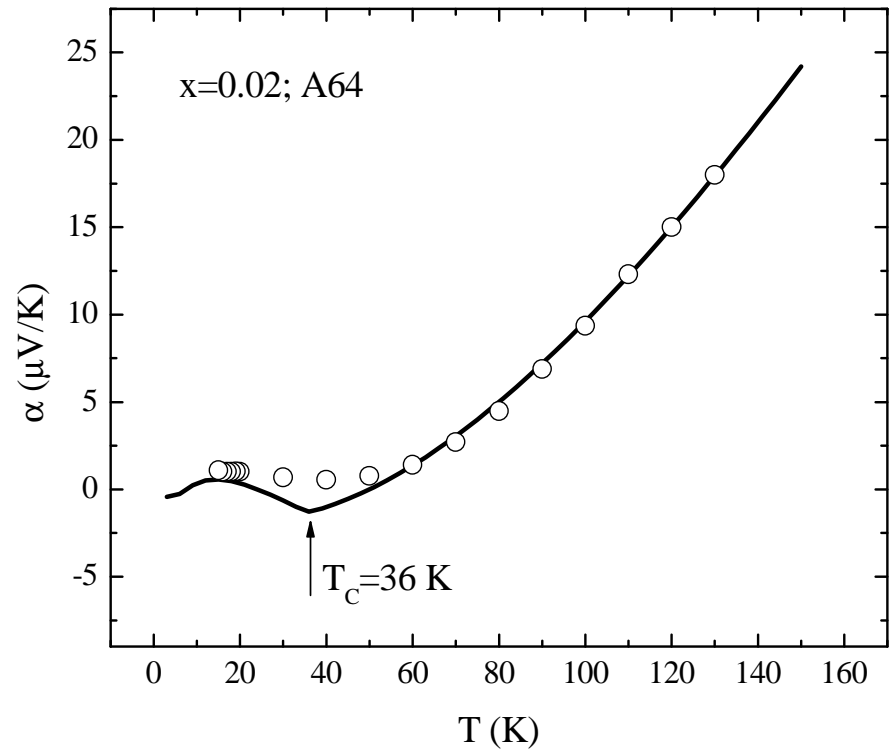
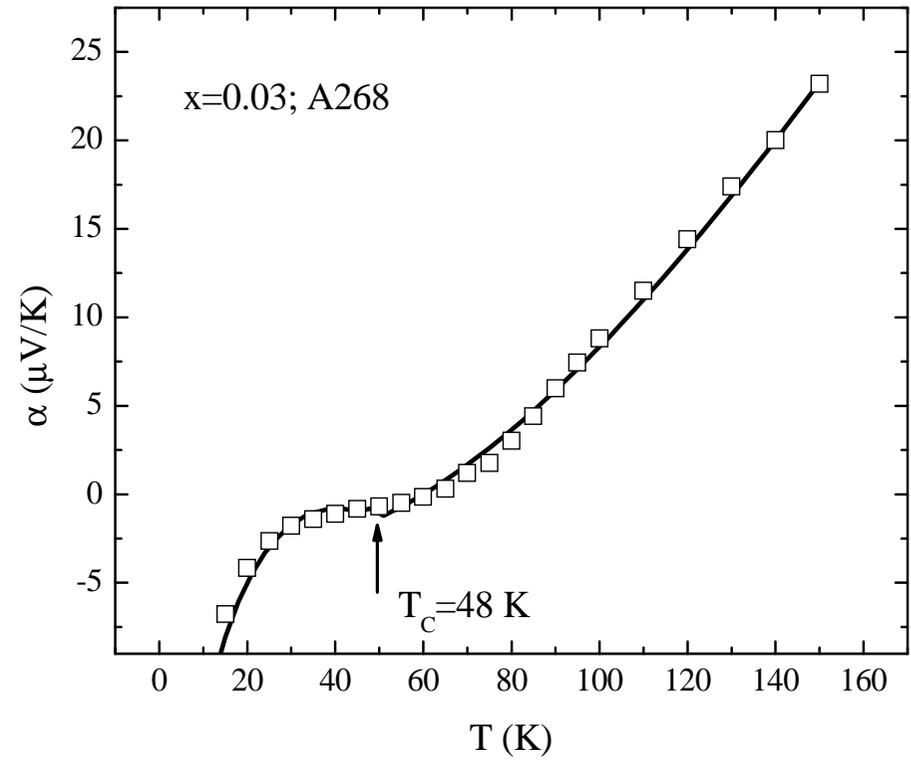

V. Osinniy et. al. Fig.7.

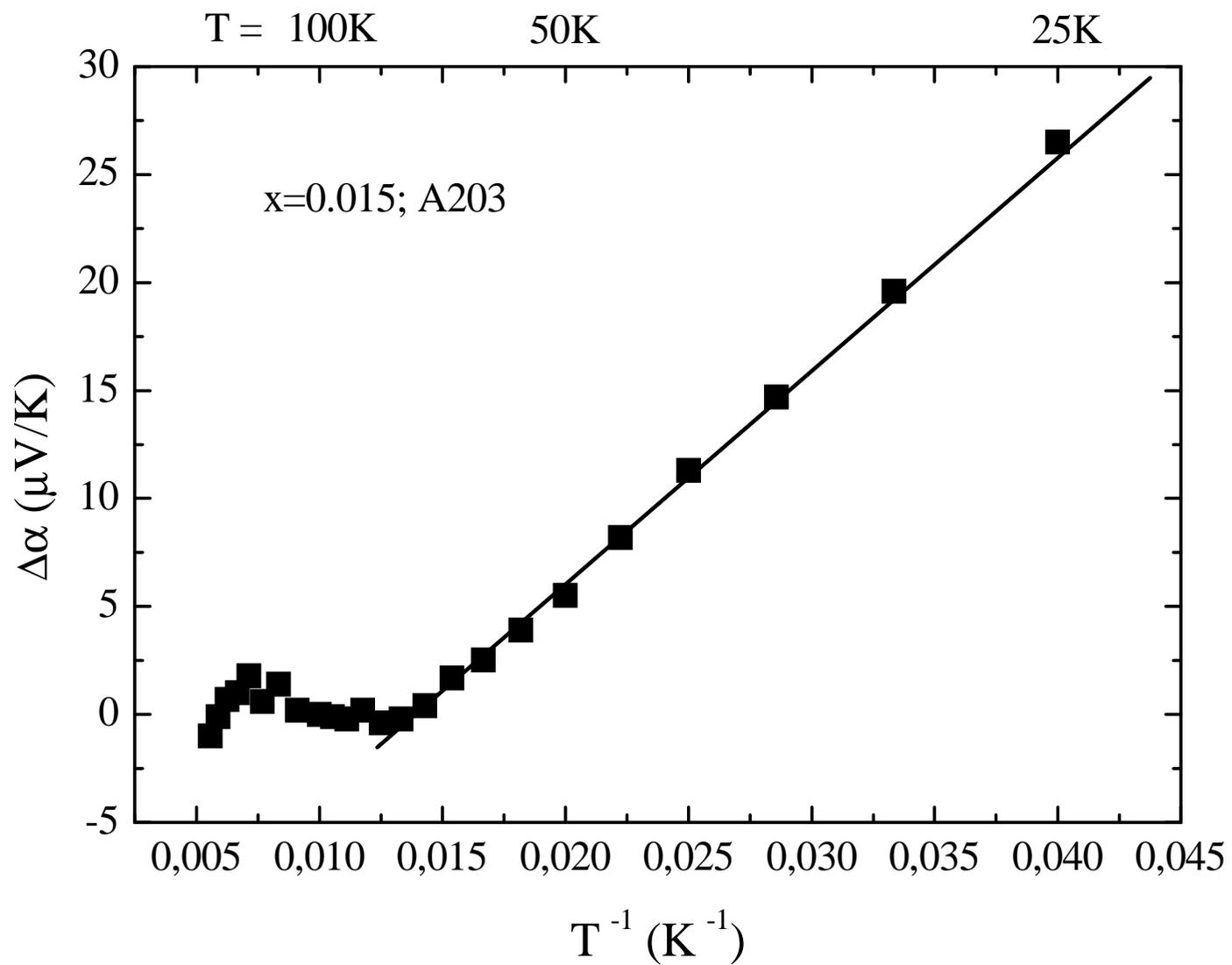

V. Osinniy et. al. Fig.8.